\documentstyle[12pt,moriond,graphics]{article}
\begin{document}
\heading{THE PHOENIX DEEP SURVEY: \\ A Deep Microjansky Radio Survey}

\author{J. Afonso $^{1}$, B. Mobasher $^{1}$, A. Hopkins $^{2}$, L. Cram $^{3}$} 
{$^{1}$ Imperial College of Science, Technology and Medicine, London, U.K.} 
{$^{2}$ Australia Telescope National Facility, Epping, NSW, Australia.} 
{$^{3}$ University of Sydney, Sydney, NSW, Australia.} 

\begin{moriondabstract}

The study of the nature of faint radio sources is of great importance since 
a significant fraction of these objects is thought to be composed of actively 
star-forming galaxies. Due to the increased sensitivity of radio telescopes, 
we are now not only able to catalogue large numbers of these sources in the 
sub-millijansky regime, but also to start the study of the nature of increasingly 
fainter microjansky sources.

This paper presents a new very deep 1.4\,GHz radio survey made as a part of the 
{\it Phoenix Deep Survey}, a project aimed to study the nature of the faintest 
radio sources. With a limiting sensitivity of 45\,$\mu$Jy, this new survey has 
allowed us to assemble a large number of sources with 1.4\,GHz flux densities 
below 100\,$\mu$Jy. The resulting source counts and the analysis of the optical 
properties of the faintest radio sources is presented.

\end{moriondabstract}

\section{Introduction}

Deep radio surveys are dominated, at sub-millijansky levels, by a population of 
actively star-forming galaxies out to $z$$\sim$1. This sub-millijansky population 
was first revealed by the changing slope of the 1.4\,GHz radio source counts below 
$\sim$5\,mJy (e.g. \cite{wind85}). Spectroscopic and photometric follow-up studies 
(\cite{benn93}, \cite{thuan92}) have revealed a population of optically blue 
galaxies, often exhibiting perturbed morphologies indicative of interactions, 
with spectra similar to those of the star-forming IRAS galaxies. Moreover, the 
study of their luminosity function has revealed that these sources have undergone 
strong luminosity evolution, possibly induced by interactions and mergers 
(\cite{hopk98}, \cite{rr93}). However, it has been hinted that at the microjansky 
level radio sources could be related to AGN activity rather than star formation 
(\cite{hamm95}) and even at higher radio levels the matter is not completely settled 
(\cite{grup99}). The assembly of large samples of these sources, down to the microjansky 
level, is essential to clarify the issue.

It is nevertheless clear that deep radio surveys can assemble large numbers of the 
most active star-forming galaxies out to $z$$\sim$1. The fact that 1.4\,GHz radiation 
can be used to determine star formation rates in star-forming galaxies (\cite{cram98}), 
with several advantages over alternative methods (insensitivity to dust absorption, 
unlike the optical and ultraviolet indicators, and superior sensitivity and astrometric 
precision over millimeter-wave and far-infrared indicators, thus allowing direct optical 
identifications) shows that deep radio surveys can lead to important studies on galaxy 
evolution and the global star-formation rate history. 

Here we present one of the faintest radio surveys made so far at 1.4\,GHz. Covering a 
wide 50$'$ diameter region, this has allowed us to assemble a large sample of sub-mJy 
and microjansky 1.4\,GHz radio sources. Using $R$-band CCD images, we also present a 
preliminary analysis of the optical counterparts of the faintest radio objects.

\section{1.4\,GHz Radio observations and source counts}

The {\it Phoenix Deep Survey} project is a collaboration between Imperial College and 
the University of Sydney, aimed to study the nature and statistical
properties of the faint radio population, through deep multiwavelength 
observations. The radio survey is carried out at 1.4\,GHz and covers an area of 3.1 
square degrees (the {\it Phoenix Deep Field} -- PDF). Previous observations resulted 
in images with a 5$\sigma$ sensitivity 
of 300\,$\mu$Jy and 100\,$\mu$Jy in a smaller 36$'$ diameter sub-region (the Phoenix 
Deep Field Sub-region -- PDFS) (\cite{hopk98}). Recently, a 1$^\circ$ diameter 
region centered on the PDFS was observed using the Australia Telescope Compact 
Array (ATCA) in its 6C configuration. Details on the data reduction are 
presented elsewhere (\cite{af99}, \cite{hopk99}). The resulting ultra-deep survey 
incorporates a total of 164 hours of observation. On the final natural weighted image 
the RMS noise increases uniformly from 9\,$\mu$Jy at the center to 25\,$\mu$Jy at a 
radius of 25$'$, the limit of the chosen area to perform source detection.

A total of 773 sources with a peak flux density greater than the local 4$\sigma$ value 
survived visual inspection. Integrated fluxes (S$_{1.4}$) range from 45\,$\mu$Jy to 
23\,mJy with 187 sources ha\-ving S$_{1.4}$$<$100\,$\mu$Jy. The flux distribution of the 
sub-mJy and microjansky sources detected is presented in Figure~1a.

The radio source counts were constructed and are presented in Figure~1b, along with data 
from other surveys (the counts from previous work on the PDF and PDFS \cite{hopk98}, a 
compilation of data from \cite{wind93} and the source counts from the 1.4\,GHz ATCA 
observations of the Hubble Deep Field South (HDF-S) \cite{norr99}). The solid and dashed 
lines represent source count predictions assuming a mixed AGN and starburst population 
(\cite{hopk98}). The best fit was achieved assuming a luminosity evolution 
($L(z) \propto (1+z)^Q$) for the starburst population at rates corresponding to $Q = 3.3$ 
(solid line). The  $Q = 2.5$ and 4.1 models are also shown (lower and upper dashed lines 
respectively). It can be seen that the models continue to provide a good fit to the observed 
source counts down to the 50\,$\mu$Jy flux level. It must be noted that these models predict 
that the majority of the sources below $\sim$200\,$\mu$Jy (down to a few microjansky) will 
be starburst galaxies. 

\vspace*{0.4truecm}
\begin{figure}[h]
\centering{
\scalebox{0.75}{
\rotatebox{90}{
\includegraphics{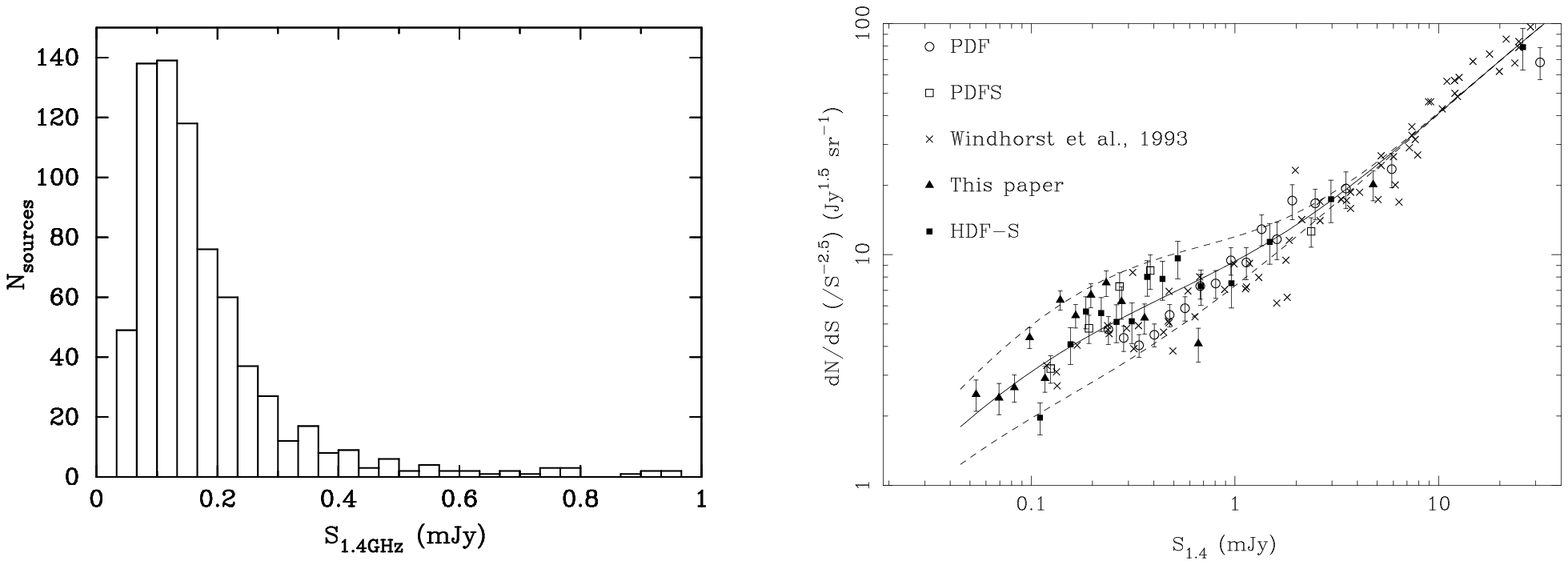}
}}}
{\footnotesize
\put(-258,43){a)}
\put(-25,42){b)}
}
\end{figure}
\vspace*{-0.1truecm}
{\footnotesize
\noindent {\bf Figure 1: a)} Radio flux distribution for the sub-mJy and microjansky 
sources detected and {\bf b)} the normalized differential 1.4\,GHz source counts for the 
present radio survey, along with data from other surveys and radio source count 
models -- see text for details.
}

\section{Optical counterparts of the microjansky sources}

The optical photometric observations of this field were carried out with the Anglo-Australian 
Telescope (AAT) and consist of prime-focus CCD observations made in the Johnson-Kron-Cousins 
$R$-band, with a completeness limit of $R$=22.5. A detailed account of the optical 
observations and data reduction is given in \cite{georg99}. 

The most probable optical association to a given radio source was chosen by searching a 
radius of 5$''$ around it and selecting the optical source with the smallest probability 
of being an acidental alignment (given the known surface density of sources as bright or 
brighter than the candidate), if less than 5\%.
Of the 773 detected radio sources, 52\%
have optical counterparts. Table 1 summarizes the analysis of the optical counterparts of 
these sources.

\vspace*{0.2truecm}
\begin{center}
{\small
{\bf Table 1:} Optical identification statistics 
}
\end{center}
\begin{center}
\begin{tabular}{lcc} \hline \hline \\[-12pt]
S$_{1.4}$ ($\mu$Jy) & ID rate (\%) & $R_{med}$ \\[2pt]
\hline \\[-10pt]
200--800 & 56 & 20.1  \\
100--200 & 55 & 20.3  \\
45--100  & 44 & 21.2  \\[2pt] \hline \hline
\end{tabular}
\end{center}
\vspace*{0.4truecm}

Both the identification rate and the median $R$ magnitude are seen to decrease for flux 
densities S$_{1.4}$$<$100\,$\mu$Jy. Also, most of the sources with radio fluxes below this 
level are identified with single optical galaxies, rather than interacting systems (see 
examples in Figure 2), unlike those above this limit. This may reflect the fact that these 
identifications are made quite close to the limit of the CCD images, or it may indicate 
that the very faint radio sources are located in galaxies different from those harboring 
slightly brighter sub-mJy radio sources. Deeper optical imaging of these sources is needed 
to clarify this issue.

\vspace*{0.5truecm}
\begin{figure}[h]
\centering{
\scalebox{0.35}{
\includegraphics*[50,180][505,625]{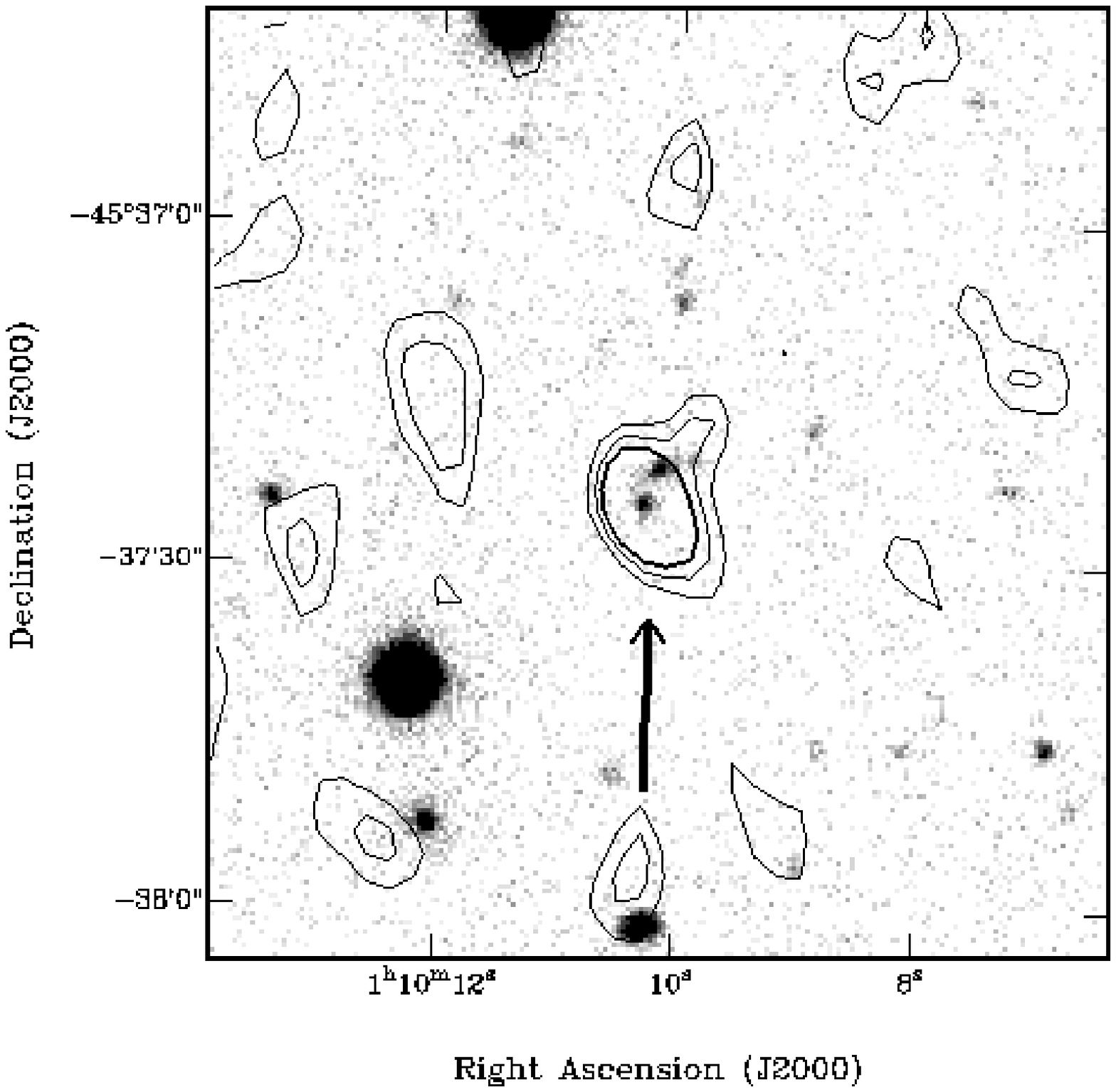}
\hspace{0.2in}
\includegraphics*[75,180][505,625]{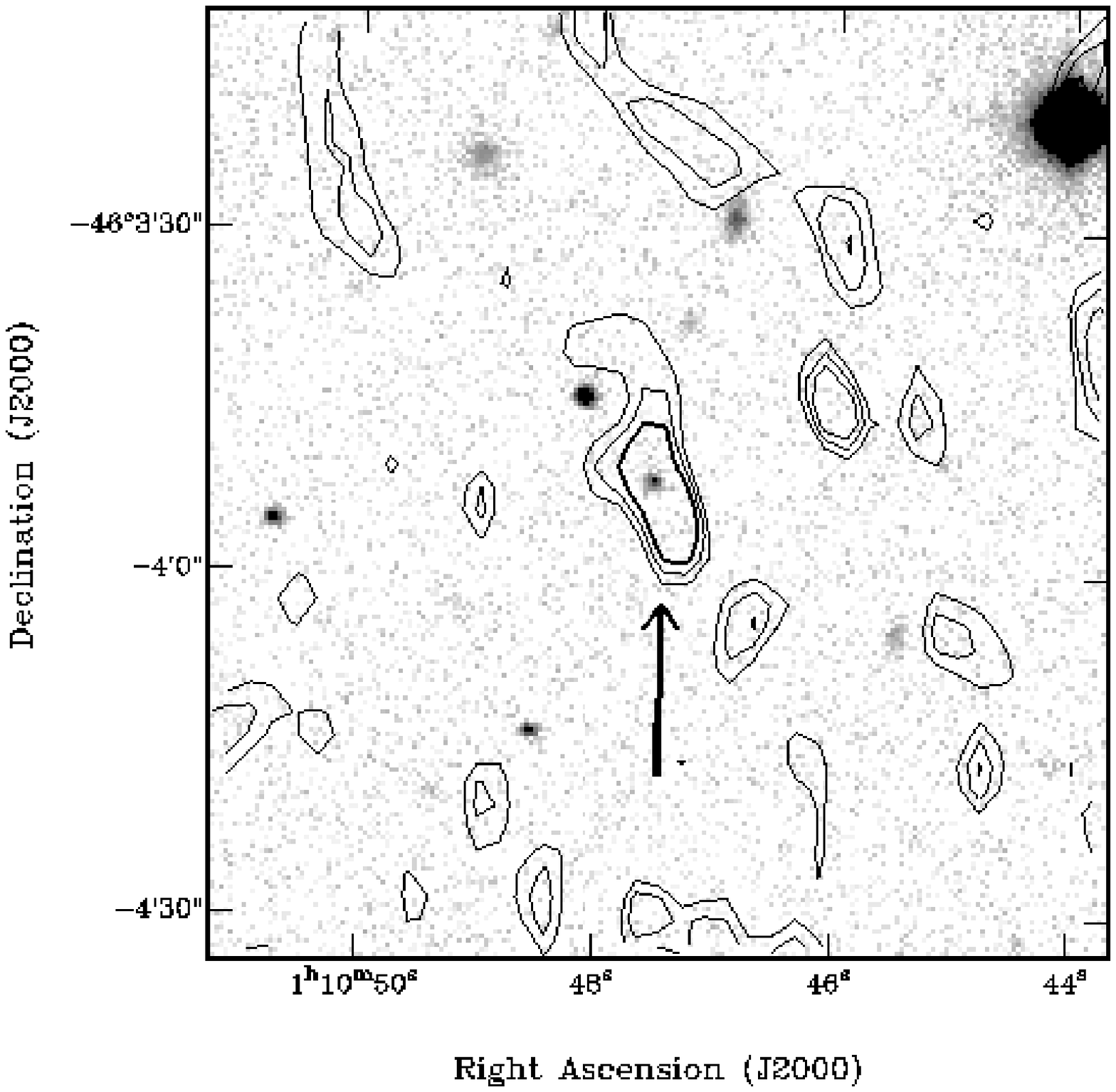}
\hspace{0.2in}
\includegraphics*[75,180][505,625]{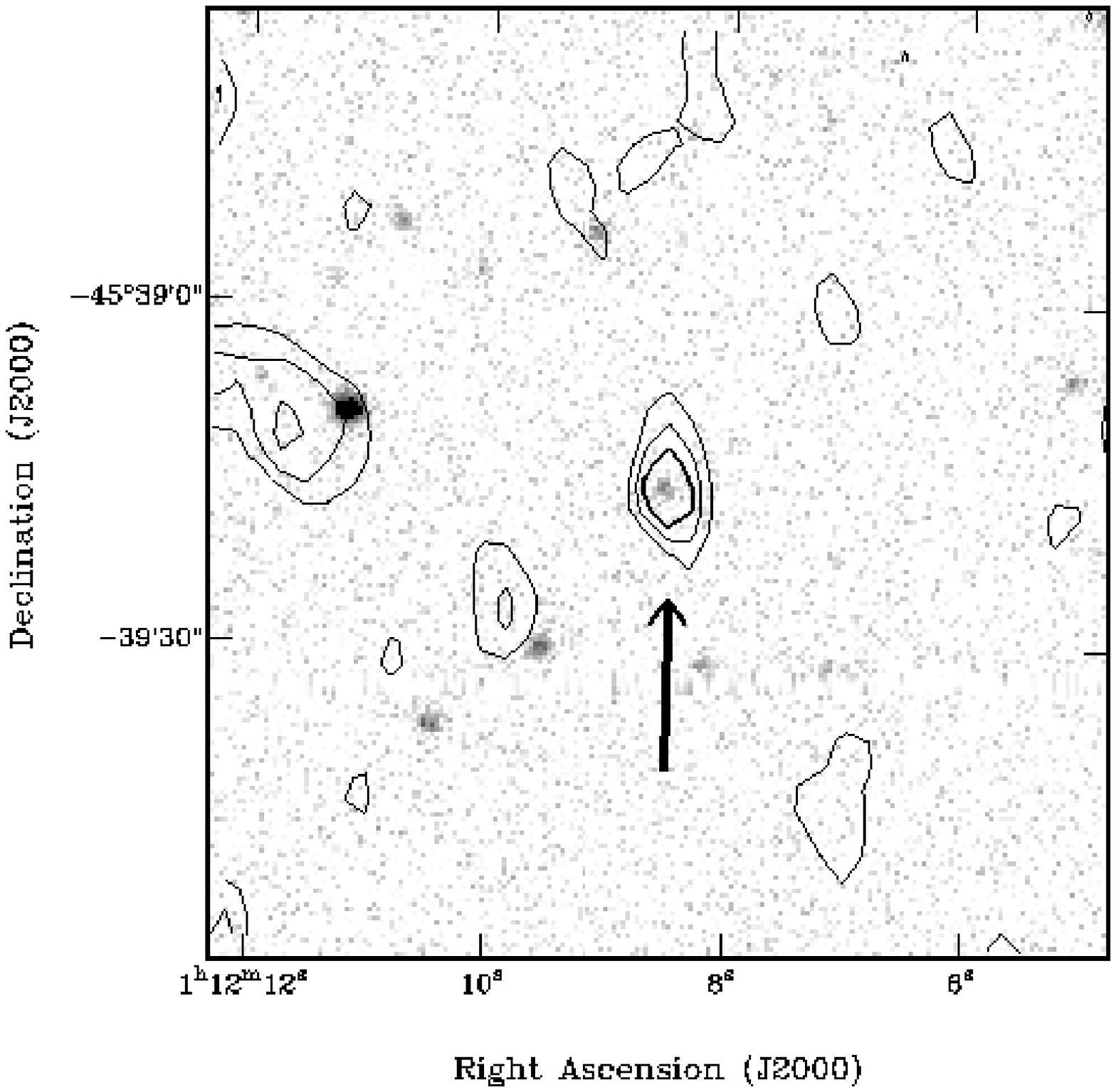}
}
}
\end{figure}
\vspace*{-0.4truecm}
{\footnotesize
\noindent {\bf Figure 2:} Examples of the optical counterparts of the microjansky sources 
detected in the present survey. The contours correspond to the 1.4\,GHz emission. The 
integrated radio fluxes of the sources indicated by the arrows are, from left to right,  
S$_{1.4}=95$, 96 and 48\,$\mu$Jy. 
}

\section{Conclusions}

A new ultra deep radio survey is presented. The sample of faint radio sources assembled, 
with S$_{1.4}$$>$45\,$\mu$Jy, is homogeneously selected and constitutes a uniform survey 
across the the entire flux density range. This sample is being used to study a large number 
of the faintest radio sources through multicolour photometry and spectroscopy.

The source counts were built and, down to the 50\,$\mu$Jy level, are seen to be consistent 
with radio source count models that imply a majority of starburst galaxies below 
$\sim$200\,$\mu$Jy. 

A significant number of sources with 1.4\,GHz fluxes below 100\,$\mu$Jy were detected and 
the analysis of their optical counterparts show that the optical magnitude of these sources 
is lower than that for the (radio) brighter sub-mJy population. Also, they are mostly 
identified with single optical galaxies, although this may reflect the relatively bright 
optical magnitude limit of the present survey.

\acknowledgements{We thank the ATCA HDF-S team for generously making their 1.4\,GHz source 
count data available. JA wishes to thank the organizers for financial support to attend the 
conference. Support from Funda\c{c}\~ao para a Ci\^encia e a Tecnologia to JA in the form of 
a scholarship is gratefully acknowledged. The Australia Telescope is funded by the Commonwealth 
of Australia for operation as a National Facility managed by CSIRO.}


%
\vfill
\end{document}